\documentclass[iop,numberedappendix]{emulateapj-rtx4}
\usepackage{graphicx,natbib,color,bm,url,times}
\usepackage{amsmath}
\usepackage{amssymb}

\newcommand{\beq}{\begin{equation}}
\newcommand{\eeq}{\end{equation}}

\begin{document}

\title{Understanding spectral states of sub-Keplerian accretion discs around compact objects
as transitions between steady states}

\author{Arunima Ajay$^1$, S  R Rajesh$^1$, and Nishant K. Singh$^2$}
\affil{
$^1$Department of Physics and Research Centre, S D College, University of Kerala, India\\
$^2$Inter-University Centre for Astronomy \& Astrophysics, Post Bag 4, Ganeshkhind,
Pune, India - 411007
}

\date{}


\begin{abstract}
We present here a simple hydrodynamic model based on a sequence of steady states of the inner
sub-Keplerian accretion disc to model its spectral states. Correlations between different hydrodynamic
steady states are studied with a goal to understand the origin of, e.g.,
the aperiodic variabilities. The plausible source of
corona/outflow close to the central compact object is shown to be a consequence of steady state transition
in the underlying accretion flow. We envisage that this
phenomenological model can give insight on the influence of environment on the inner sub-Keplerian
accretion disc.
\end{abstract}
\keywords{accretion, accretion discs --- hydrodynamics --- X-rays: binaries, AGN}
\email{srr@sdcollege.in, nishant@iucaa.in}

\section{Introduction}
Accretion discs around compact objects, such as, black holes and neutron stars,
are powerful sources of energy in the Universe \citep{FKR02}. Accretion systems vary from
ultra luminous X-ray sources, e.g., SS433, to under luminous AGNs/quasars, e.g.,
Sgr A$^\ast$ \citep{MFK79, MM07}. Ever since the foundations of the theory of steady state accretion
flow was established, considerable progress has happened on both observational and
theoretical fronts. In current picture of the accretion physics, there is a
sub-Keplerian accretion disc close to the central compact object, fuelled by a
much larger outer Keplerian accretion disc. It is thought that within
the sub-Keplerian disc itself,
there could be different flow domains which are characterised by the parameters of
the fluid, such as, viscosity, magnetic field and the cooling mechanism.
In such a standard accretion theory, the state of the disc
is defined by the steady state values of mass accretion rate, viscosity and the
efficiency of the radiative cooling mechanism
\citep[see the review,][]{YN14}.

An accretion disc/flow domain is
considered to be in a specific state if these steady state parameters do not change
appreciably for a considerable amount of time.
However, relatively new observations of aperiodic variabilities	\citep{MV97,HB05,RM06},
and temporal variabilities in radiative output are now understood in terms of
the transitions of the inner sub-Keplerian accretion disc among different spectral
states \citep{TH72,WNP01}.

Although the accretion systems vary in their sizes, environments etc,
one nevertheless expects certain universality, as all of these systems
essentially involve accretion of hot plasma with similar underlying dynamics
of the flow around the central compact object.
Taking into account the effects of the environments in which the accretion
systems reside appears to be essential to better understand the
associated radiative outputs \citep{GCM01, CC20}.
These studies therefore suggest that an attempt to build a theoretical model
to accommodate the influence of the environment on the states of an accretion disc
is essential.

Spectral transitions that are observed in the accretion systems, such as X-ray binaries,
are identified as Soft-High, Hard-Low, and several other intermediate
states.
During the aperiodic transitions, many of these accretion systems remain in the
intermediate states for long duration \citep{EPP75, SSW79, GMM06, BJ10}.
In many cases it is also observed that the system returns back to the original state,
implying that the system retains some ``memory''. These observed spectral state
transitions are thought to originate from the
inner part of the accretion disc close to the compact object, leading to the
assumption that the different spectral states correspond to different hydrodynamic steady states
of sub-Keplerian accretion disc.

In this work, we propose a simple phenomenological model where the
sub-Keplerian accretion disc
is assumed to undergo a transition between two hydrodynamic steady states.
This is expected to give rise to distinct spectral states that are observed.
First, we ask the following question: in the transition between two steady
states, how much information about the first (initial) state is retained in the
second (intermediate) state? This question is inspired by the fact
that observationally we find that during the aperiodic spectral state transitions,
the accretion system eventually comes back to the original initial state. Hence,
the intermediate state must have some amount of memory/information about the
initial state \citep{TH72, CEQ76, WNP01, GMM06}.
Often the intermediate states are followed by subsidiary components such as
corona or outflow; see, for example, the case of GRS 1915+105 as discussed in \cite{FG99, MR94}. Our model naturally incorporates the possibility of source of corona or outflow
during steady state transitions for a suitable range of flow parameters.

The paper is organized as follows. In \S 2, we present our model with the relevant hydrodynamic
equations and the series solutions. The description of steady states along with the conditions
for transitions are then presented. In \S 3, the self-similar linear first order solutions for a
self-similar background accretion disc are discussed. In \S 4, the fully nonlinear first order conservation
equations for a self-similar background accretion disc are introduced.
The numerical results of different types of plausible transitions are discussed in \S 5. The summary of the work and its relevance in astrophysical observations are discussed
in \S 6.

\section{Description of the model}
\label{des}

We assume the standard $\alpha$ prescription for the viscosity which varies in the
range $10^{-3} < \alpha < 1$ \citep{SS73, PR72, NT73}. While protons carry much of the
turbulent energy, electrons are mainly responsible for the radiative cooling where
non-thermal Bremsstrahlung is the dominant mechanism in the sub-Keplerian part of
the disc. Wherever the magnetic fields are significant, cooling is then dominated
by synchrotron radiation which is further boosted by the inverse Comptonization
process \citep{NY95, MC05, RM10}.
Here we parameterise the energy advected by the radial flow using the
advective factor $f$ which is defined as the fraction of viscously
dissipated energy advected by the flow. The value of $f$ is unity in an
advection dominated accretion flow, ADAF \citep{NY94, NY95}, whereas it is zero for an outer cool
Keplerian disc where the radiative cooling is highly efficient. 
It lies in the range $0 < f < 1$ for a two-temperature
sub-Keplerian accretion disc \citep{RM10,RBM10}, and, in a sense, effectively determines the
efficiency of the cooling mechanism, details of which can be worked out using energy equation for electron flow.
We assume $f$ to be a constant throughout the inner sub-Keplerian accretion disc
in a given steady state configuration.

A scheme of our model may be presented as follows. Mass accretion rate is the
primary parameter determining the hydrodynamic steady state of the disc.
Outer environment such as the Keplerian part of the disc communicates with
the inner sub-Keplerian disc via variation in the mass accretion rate, which
if finite, may lead to transitions to energetically excited steady states, such
as ADAF or convection dominated accretion flow (CDAF).
In these energetic states, enough free energy is available for the system to
organise itself to manifest different phenomena, such as outflow, creation of corona,
hard X-ray emission etc.
In our simplistic model, the origin of these phenomena are modelled as a
consequence of transition of the system between different energetic steady states.
Source of corona/outflow is modelled in terms of mass flow rate out of inner disc.
We define a smallness parameter $\varepsilon$ which is the ratio of the source
mass rate of corona/outflow to the mass rate of the disc.

Let us consider that $\alpha_{0}$ and $f_{0}$ be the viscosity parameter and
advective factor, respectively, of the energetic, e.g., the ADAF state.
Facilitated by variation in mass accretion rate, the disc could undergo a
transition to a new hydrodynamically steady state, say, the intermediate state
with viscosity and advective factor being $\alpha$ and $f$, respectively.
As mentioned above, the transition from the state with parameters
$\alpha_{0}$ and $f_{0}$ to a state with parameters $\alpha$ and $f$ is due to
the intrinsic tendency to stabilise the system, and this may lead to corona/outflow
which is parametrised  by $\varepsilon$.
With such an overview as outlined above, we now study the hydrodynamic conservation
equations of the accretion disc in the following. 

\subsection{Conservation equations}
We write the conservation equations for accretion disc in cylindrical polar
coordinates. Here we study the steady states of an axisymmetric accretion disc,
therefore, the hydrodynamic equations are independent of time and angular
coordinate. Then the standard vertically integrated hydrodynamic equations of
the accretion flow with viscosity $\alpha$, advective factor $f$, and total mass
accretion rate $\dot{M}$ are: \\

(i) conservation of mass,
\begin{equation}
4 \pi r \bar{\rho} h v= \dot{M}
\label{ozr1}
\end{equation}

(ii) conservation of radial momentum,
\begin{equation}
\bar{\rho} v \frac{\partial  v }{\partial r}  - \bar{\rho} \Omega^2r =  - \frac{\partial  (\bar{\rho} c_s^2) }{\partial r}- \frac {\bar{\rho}}{r^2}
\label{ozr2}
\end{equation}

(iii) conservation of angular momentum,
\begin{equation}
\bar{\rho} v  h \frac{\partial  (\Omega r^2) }{\partial r} = \frac {1}{r} \frac{\partial  (r^2  \mathcal{T}^{r \theta} h) }{\partial r}
\label{ozr3}
\end{equation}

(iv) conservation of energy,
\begin{equation}
\frac{v h}{(\gamma-1)} \left(  \frac{\partial  (\bar{\rho} c_s^2) }{\partial r} - \gamma  c_s^2 \frac{\partial  \bar{\rho} }{\partial r}\right) = f \mathcal{T}^{r \theta} h  r \frac{\partial  \Omega }{\partial r}\,,
\label{ozr4}
\end{equation} 

\noindent
where, the radial velocity $v$, the angular velocity $\Omega$ and the square of the sound speed $c_s^{2}$ are functions of radial co-ordinate $r$. The scale height of the disc
is denoted by $h$. The density $\rho$ and the total pressure $p$ are functions of radial co-ordinate $r$ and the vertical co-ordinate $z$. The total pressure $p$ is the sum of the partial pressures such as gas pressure, radiation pressure and magnetic pressure.  The over bar symbol means that the quantity is vertically averaged. Pressure is related to density and sound speed through equation of state given by the expression 
\begin{equation}
\bar{p} =  \bar{\rho} c_s^{2}
\end{equation}
and $\gamma$ is the ratio of molar specific heats at constant pressure and volume. For pure gas $\gamma$ is $5/3$ and for pure radiation $\gamma$ is $4/3$. In general the matter is a mixture of gas and radiation, therefore the value of $\gamma$ is between $4/3$ and $5/3$. Since we perform our analysis at sufficiently far away from the central compact object, such as the black hole, it is enough to consider the Newtonian gravity. We scale the mass, length and speed in the units of $M$, $GM/\emph{c}^2$ and $\emph{c}$, respectively, where $M$ is the mass of the central compact object expressed in units of solar mass in this paper, $G$ is the gravitational constant and $\emph{c}$ is the speed of light in free space. In these units the
magnitude of the gravitational force is $F= 1/r^2$. The stress tensor is expressed in the Mixed - Stress form
\citep{NY94}.
The vertically integrated stress tensor is
\begin{equation}
\mathcal{T}^{r \theta} h = \alpha\bar{\rho} h c_s^2 \frac{\partial  \Omega }{\partial r} r^{5/2}
\end{equation}

\subsection{Series expansion of the flow variables}
As defined in \S~\ref{des}, the smallness parameter $\varepsilon$ is the ratio of the mass rate of the plausible sources of corona/outflow to the mass rate of the accretion disc. Out of the total mass accretion rate $\dot{M}$, if $\dot{M}_0$ represents the mass rate of the background disc then 
\begin{equation}
\varepsilon \, = \,  \frac{\dot{M} \, - \, \dot{M}_{0}}{\dot{M}_{0}} 
\label{exp1}
\end{equation}
where $\varepsilon\ll1$ for the class of transitions discussed in this work. 
We express the fluid variables of the intermediate state as polynomial series in $\varepsilon$.
\begin{equation}
\rho (r,z) \, =\, \sum\limits_{n=0}^\infty \rho_{n}(r,z) \, \, \varepsilon^n
\, \,  , \, \, 
c_s^2 (r)\, = \, \sum\limits_{n=0}^\infty c^2_{sn}(r) \, \, \varepsilon^n
\label{vaf2}
\end{equation}
\begin{equation}
v(r) \, = \, \sum\limits_{n=0}^\infty v_{n}(r) \, \,  \varepsilon^n
\, \, , \, \, 
\Omega (r) \, =\, \sum\limits_{n=0}^\infty \Omega_{n}(r) \, \, \varepsilon^n
\label{vaf4}
\end{equation}
where $n \, = \, 0,  \, 1, \, 2, \, 3 \ldots$. The total density is the sum of the functions $\rho_{0}$, $\rho_{1}$ etc. All these density functions are symmetric with respect to the $z$ coordinate. The scale heights of these functions are $h_{0}$, $h_{1}$ etc and the total height of the disc is $h$. Since $h > h_{n}$, we can write
\begin{equation}
\bar{\rho} h = \bar{\rho_{0}}  h_{0} + \varepsilon \bar{\rho_{1}} h_{1} + ...
\end{equation}
The vertical equilibrium equation for the accretion disc is given by 
\begin{equation}
-\frac{\partial  p }{\partial z} + \rho g^z = 0
\label{fce4}
\end{equation}
where $g^z \, = -F z/r\, $. We substitute the expressions
of density, $\rho$ and square of sound speed, $c_s^{2}$ (Eqn.\ref{vaf2}) in the above equation and equate the same powers of $\varepsilon$. We calculate the scale heights $h_{0}$, $h_{1}$ etc by vertically integrating these equations.  For simplicity we avoid the over bar symbol for density functions such that now onwards $\rho_{0}$, $\rho_{1}$ etc mean height averaged quantities. Thus we have the expressions: 
\begin{equation}
h_{0} = c_{s0} r^{1/2} F^{-1/2} \; \;
\mbox{and} \;\;
h_{1} = h_{0} \phi^{1/2} 
\label{hexp1}
\end{equation}
where
\begin{equation}
\frac{h_{1}}{h_{0}} = h_{1,0} = \phi^{1/2} = \left(1+ \frac{c_{s1}^2 \rho_{0}}{c_{s0}^2 \rho_{1}}\right)^{1/2}
\end{equation}

\subsection{Correlation among the flow parameters}

In order to find the conservation equations of the different orders of flow variables of the intermediate steady state (ie; $\rho_n, \, v_n, \, \Omega_n, cs_n^2$ where $n=0, \, 1 \, ....$), we need to establish physically acceptable correlation between the set of parameters of the energetic initial steady state and the set of parameters of the intermediate steady state.

We express the total mass accretion rate as: 
\begin{equation}
\dot{M} =   \sum\limits_{n=0}^\infty  \dot{M}_{n} \, \,  \varepsilon^n
\label{exp2}
\end{equation}
Combining expressions ($\ref{exp1}$) and  ($\ref{exp2}$) we have: 
\begin{equation}
\dot{M_{0}} = \dot{M_{1}} + \varepsilon \,  \sum\limits_{n=0}^\infty  \dot{M}_{n+2}  \, \, \varepsilon^n
\end{equation}
where $\dot{M}_{n}$'s are inherently statistical in nature. They determine the amount of matter contained in the corresponding order of flow. Considering  all $\dot{M}_{n}$'s are of similar magnitude, we see that $\dot{M}_{1}$ is more close to $\dot{M}_{0}$ at least by a factor of $\varepsilon$ than the combined effect of all other $\dot{M}_{n}$. Physically this means that most of the mass related to the plausible source of corona/outflow is contained in the first order density,
i.e., $\dot{M}_{1} \,= \,  \dot{M}_{0} \delta$ where $\delta \rightarrow 1$. 

The hydrodynamic variables of the intermediate steady state as well as the total mass accretion rate are expressed as polynomial series in the parameter $\varepsilon$ (Eqn.\ref{vaf2}, \ref{vaf4} and \ref{exp2}). Therefore we may expect to express the correlation between the intermediate steady state viscosity parameter $\alpha$ or the intermediate steady state advective factor $f$ to the corresponding initial steady state parameters $\alpha_{0}$ or $f_{0}$ as polynomial series in $\varepsilon$. If we express $\alpha$ or $f$ as series in increasing power of $\varepsilon$ then we will arrive at the following unphysical conclusions: (1) The steady state hydrodynamic equations of the zeroth order flow variables of the intermediate state will be completely independent of the first order and other higher order flow variables. But the steady state hydrodynamic equations of the first order flow variables will depend on the the zeroth order flow variables, and independent of the second order and other higher order flow variables and so on. Thus the fundamental principal of action and reaction is not captured. (2) By definition $\alpha_{0}$ and $\alpha$ have values between zero and one. Therefore if we express $\alpha$ as series in increasing power of $\varepsilon$, then for the term $\varepsilon \alpha_{1}$ to have any significant value, the magnitude of $\alpha_{1}$ should be close to one (since $\varepsilon \ll 1$). If we continue like this, magnitudes of the higher order parameters $\alpha_{2}$, $\alpha_{3}$.. will become greater than unity and the values of these parameters will not converge as order increases.  Similar argument is valid for $f_{0}$, $f$ and higher order advective factors $f_1$, $f_2$....

We express the viscosity parameter and advective factor of the intermediate steady state as: 
\begin{equation}
\alpha=\sum\limits_{n=0}^\infty \alpha_{n} \, \, \varepsilon^{-n} \, \, , \, \, f=\sum\limits_{n=0}^\infty f_{n} \, \, \varepsilon^{-n}
\label{vissr}
\end{equation}
The above expressions are the appropriate representation of $\alpha$ and $f$ because of the following reasons: (1) The hydrodynamic equations of flow variables of different orders are coupled. Therefore the action reaction principle is maintained. (2) The values of $\alpha_0$ and $\alpha$ are between zero and one. Since $\varepsilon \ll 1$, $\varepsilon^{-n}$ increases rapidly as $n$ increases. Therefore the magnitudes of the set of numbers, the higher order viscosity parameters $\alpha_1$, $\alpha_2$, $\alpha_3$...are less than one and decrease rapidly as the order $n$ increases. Similar explanation is valid for higher order advective factors. (3) For given values of $\alpha_0$, $\alpha$ and $\varepsilon$, we can choose the set of numbers $\alpha_{1}$, $\alpha_{2}$, $\alpha_{3}\,\ldots$. Similarly for given values of $f_0$, $f$ and $\varepsilon$, we can choose the set of numbers $f_{1}$, $f_{2}$, $f_{3}\,\ldots$.
By definition $\varepsilon$ characterises the amount of matter lost by the disc due to state transition. Therefore if there is no state transition that is
$\alpha \rightarrow  \alpha_{0}$ and $f \rightarrow f_{0}$ then $\varepsilon \rightarrow 0$. For this particular combination of $\alpha_0$, $\alpha$ and $\varepsilon$, the only possible set of physically acceptable higher order viscosity parameters is $\alpha_{1} \, , \, \alpha_{2}\, , \, \alpha_{3}.... =  0$. Similarly for this particular combination of $f_0$, $f$ and $\varepsilon$, the only possible set of physically acceptable higher order advective factors is $f_{1} \, , \, f_{2}\, , \, f_{3}.... =  0$. 
In other words, $\alpha_n$ and $f_n$ decrease faster than the increase in $\varepsilon^{-n}$ for $n>0$ in the series expansion of Eqn.~\ref{vissr}.
Therefore in Eqn. \ref{vissr}
there is no issue of divergence as $\varepsilon \,\rightarrow \, 0$.

\subsection{The stress tensor and the energy advection}
For any significant amount of matter loss from the disc, $10^{-4}<\varepsilon<10^{-1}$. Therefore from the converging series Eqn.\ref{vissr},
we can easily conclude that $\alpha_{2}, \, \alpha_{3}.....\,\ll\, \alpha_{1}$ and $f_{2}, \, f_{3}....\, \ll\, f_{1}$.  As a consequence, without any loss of information we can set $\alpha_{2}, \, \alpha_{3} ...$ and $f_{2} \, , \, f_{3}..$ to zero. Since most of the matter of the plausible corona/outflow is contained in the first order flow, we neglect the terms with higher order ($n>1$) flow variables in the expressions for zeroth order and first order stress tensors and energy advection. The vertically integrated stress tensor of the intermediate state of the accretion disc can then be written as,

\begin{equation}
\mathcal{T}^{r \theta} h =G_{0} \, +\, \varepsilon G_{1} + \,\ldots \,,
\label{strsexp}
\end{equation}
where,
\begin{eqnarray}
&&G_{0} = \alpha D_{0} + \alpha_{1} D_{1}\;;\;\;
G_{1} = \alpha_{0} D_{1} + \alpha_{1} D_{2} \nonumber \\
&&G_{2} = \alpha_{0} D_{2} \;;\;\;
D_{0} = \rho _{0} h_{0} c_{s0}^2 \frac{\partial  \Omega_{0} }{\partial r} r^{5/2} \nonumber \\
&&D_{1} = \left(\rho_{0}h_{0} c_{s1}^2 \frac{\partial  \Omega_{0} }{\partial r} + 
\rho_{1} h_{1}c_{s0}^2 \frac{\partial  \Omega_{0} }{\partial r} + 
\rho_{0} h_{0}c_{s0}^2 \frac{\partial  \Omega_{1} }{\partial r} \right) r^{5/2} \nonumber \\ 
&&D_{2} = \left(\rho_{1}h_{1} c_{s1}^2 \frac{\partial  \Omega_{0} }{\partial r} +
\rho_{1} h_{1}c_{s0}^2 \frac{\partial  \Omega_{1} }{\partial r} +
\rho_{0} h_{0}c_{s1}^2 \frac{\partial  \Omega_{1} }{\partial r}\right) r^{5/2} \nonumber
\end{eqnarray}
The total amount of the turbulent energy advected by the intermediate state of the accretion disc is given by the expression, 
\begin{equation}
f \mathcal{T}^{r \theta} h r\frac{\partial \Omega}{\partial r} \, = \, f F_{0} \, + \, f_{1} F_{1} \, + \, \varepsilon ( f_{0} F_{1} + f_{1} F_{2} ) + \ldots \,,
\label{enadvexp}
\end{equation}
where,
\begin{eqnarray}
&&F_{0} = G_{0} r \frac{\partial  \Omega _{0}}{\partial r} \;; \;\;
F_{1} = G_{0} r \frac{\partial  \Omega _{1}}{\partial r} + G_{1} r \frac{\partial  \Omega _{0}}{\partial r} \nonumber \\
&&F_{2} =  G_{1} r \frac{\partial  \Omega _{1}}{\partial r} + G_{2} r \frac{\partial  \Omega _{0}}{\partial r} \nonumber
\end{eqnarray}

\subsection{Zeroth order conservation equations and the background accretion disc}
Substituting expressions of the flow variables (Eqn.\ref{vaf2}-\ref{vaf4}), vertically integrated stress tensor (Eqn.\ref{strsexp}), vertically integrated energy advection (Eqn.\ref{enadvexp}) and total mass accretion rate (Eqn.\ref{exp2}) in Eqn.\ref{ozr1} to \ref{ozr4}, and equating the zeroth power of $\varepsilon$, we obtain the conservation equations of the zeroth order flow variables of the intermediate steady state of the accretion disc.
\begin{equation}
4 \pi r \rho_{0}h_{0} v_{0} = \dot{M_{0}}
\label{zr1}
\end{equation}
\begin{equation}
\rho_{0} v_{0} \frac{\partial  v_{0} }{\partial r}  - \rho_{0}\Omega_{0}^2r =  - \frac{\partial  (\rho_{0} c_{s0}^2) }{\partial r}- \frac {\rho_{0}}{r^2}
\label{zr2}
\end{equation}
\begin{equation}
\rho_{0} v_{0} h_{0}\frac{\partial  (\Omega_{0} r^2) }{\partial r} = \frac {1}{r} \frac{\partial  }{\partial r}  \left[ r^{2}  \left( \alpha D_{0} \, + \, \alpha_{1} D_{1} \right) \right]
\label{zr3}
\end{equation}
\begin{equation}
\frac{v_{0}h_{0}}{(\gamma-1)} \left[  \frac{\partial  (\rho_{0} c_{s0}^2) }{\partial r} - \gamma  cs_{s0}^2 \frac{\partial  \rho_{0} }{\partial r}\right] = \left( f F_{0} \, + \, f_{1}F_{1} \right) r \frac{\partial  \Omega _{0}}{\partial r}
\label{zr4}
\end{equation} 
In the limiting case of $\varepsilon \rightarrow 0$ such that $\alpha_1$, $f_1 =0$ we have $\alpha=\alpha_0$, $f=f_0$. Then the above set of hydrodynamic equations of the zeroth order flow variables become same as that of the hydrodynamic equations of the initial state. 

We define two functions $\kappa_{1}$ and $\kappa_{2}$ such that 
\begin{equation}
\kappa_{1} \, = \, \left| {\frac{\alpha_{1} D_{1}}{\alpha D_{0}}}   \right|  \, \,  , \, \, \kappa_{2} \, = \,  \left |{\frac{\, f_{1} F_{1}}{f F_{0}}}  \right |
\label{expkp}
 \end{equation}
In the previous subsection \S 2.4 we saw that the higher order viscosity parameters \& higher order advective factors (for $n>1$)  are very small compared to the zeroth order \& first order coefficients and hence they are set to zero. Therefore we can write $\alpha_{1}= \left( \alpha - \alpha_{0} \right) \varepsilon$ and $f_{1} = \left( f - f_{0} \right) \varepsilon$.  Substituting these expressions in Eqn.\ref{expkp}, we find that $\kappa_{1}$ and $\kappa_{2}$ are proportional to $\varepsilon$. Hence it is reasonable to assume the condition
\begin{equation}
\kappa_{1} \, \, , \, \,  \kappa_{2} \, \, \lesssim \, \, 10^{-1}
\label{apprx}
\end{equation}
Therefore we neglect the terms with parameters $\alpha_{1}$ \& $f_{1}$ in the expressions for stress tensor and energy advection in the zeroth order hydrodynamic equations. With this approximation, the set of conservation equations of the zeroth order flow variables of the intermediate steady state has the form of the set of standard accretion disc equations with mass accretion rate $\dot{M_0}$, viscosity $\alpha$ and advective factor $f$. The mass, momentum and energy of plausible source corona/outflow will be governed by the first order flow. The corresponding hydrodynamic equations are discussed in the following subsection. 

The aim of our model is to understand the states and transitions of a sub-Keplerian accretion disc which is sufficiently away from the central compact object such that general relativistic effects can be neglected. Also most of the radiative flux which shows variabilities in high energy band comes from the inner accretion disc at radius $r<10^{3}$. Therefore for the region which we study, the self-similar solution \citep{NY94} can faithfully represent the background accretion disc, which is: 
\begin{equation}
\rho_{0} = A r^{-\frac{3}{2}} \, ; \, c_{s0}^2 = B r^{-1} \, ; \, v_{0} = C r^{-\frac{1}{2}} \, ; \, \Omega_{0} = D r^{-\frac{3}{2}}
\label{solfn}
\end{equation}
where the numbers $A$, $B$, $C$ and $D$ are determined by the parameters $\dot{M}_{0}$, $\alpha$ and $f$. 

\subsection{First order conservation equations and transition}
Similar to \S2.5, by equating the first power of $\varepsilon$ in Eqn.\ref{ozr1} to \ref{ozr4}, we obtain the conservation equations of the first order flow variables of the intermediate steady state of the accretion disc. For convenience the set of first order conservation equations are given in Appendix-I, Eqns.\ref{fr1}-\ref{fr4}. The first order flow variables represent the change in density, velocity and energy of the intermediate steady state of the accretion flow due to the component of the total mass accretion rate, $\varepsilon \dot{M_{0}}$. Similar to the zeroth order equations, we neglect the terms with first
order parameters in the expressions for the first order stress tensor and the first order energy advection. The first order continuity equation (Eqn.\ref{fr1}) takes care of the mass influx which could plausibly be diverted out of the disc. The corresponding changes in radial velocity and angular velocity are taken care of by the first order radial momentum equation (Eqn.\ref{fr2}) and the first order angular momentum equation (Eqn.\ref{fr3}), respectively. The left hand side of the first order energy equation (Eqn.\ref{fr4}) has two parts.  The first part is the energy due to the interaction between the zeroth order flow variables and the first order flow variables, which is advected by the zeroth order radial velocity. The second part is the energy due to the zeroth order flow variables, which is advected by the first order radial velocity. 

For an accretion disc of total mass accretion rate $\dot{M}$, which has undergone transition to an intermediate steady state of parameters $\alpha$ and $f$, up to the first order approximation the total hydrodynamic solution is:
\begin{equation}
\rho=\rho_0+\varepsilon \rho_1, \,v=v_0+\varepsilon v_1, \,\Omega=\Omega_0+\varepsilon \Omega_1, \,c_s^2=c_{s0}^2 + \varepsilon c^2_{s1} 
\label{tsln}
\end{equation}
The zeroth order solution (Eqn.\ref{solfn}) depends on the parameters of the current/intermediate state of the accretion disc ($\alpha$, $f$). Unlike the zeroth order solution, the first order solution depends on the parameters of both the initial and the intermediate states ($\alpha_0$, $f_0$ and $\alpha$, $f$). Thus the memory of the previous state is embedded in the first order solution. This is consistent with the requirement of the model discussed in \S1. The fundamental assumption while suggesting a state transition of an accretion disc from a steady state with parameters $\alpha_{0}$ and $f_{0}$ to a plausible steady state with parameters $\alpha$ and $f$ is that the transition is able to support a first order solution. For a given transition if there exists no first order solution connecting the outer boundary to the inner boundary of the accretion flow domain, then that will violate the conservation of mass, momentum and energy. Hence such a transition is prohibited.

When $\varepsilon \rightarrow  0$, the first order parameters
$\alpha_1, f_1  =  0$ such that $\alpha=\alpha_0$, $f=f_0$ and $\dot{M_0}=\dot{M}$. Then the hydrodynamic solution $\rho, v, \Omega$ and $c_s^2$ becomes the steady state solution of accretion disc of viscosity parameter $\alpha_0$, advective factor $f_0$ and mass accretion rate $\dot{M}$. Thus in the limiting cases of $\varepsilon \rightarrow 0$ we regain the initial steady state accretion disc with parameters $\alpha_0$ and $f_0$. If we symbolically represent the steady state of the accretion disc as a set of viscosity, advective factor and mass accretion rate then the steady state transition of the accretion disc can be represented as: 
\begin{equation}
(\alpha_0, \, f_0 , \, \dot{M}) \longrightarrow (\alpha , \, f, \, \dot{M_0}) +  (\alpha , \, \alpha_0 , \, f , \, f_0 , \, \varepsilon \dot{M_0})
\end{equation}
Where the last term symbolically represents the first order solution. The theory developed in \S2 is general and is independent of the type of accretion solutions. However in this work we study the transition between two self-similar background flows (zeroth order solution, Eqn.\ref{solfn}). In the following sections we discuss analytical as well as numerical solutions of the first order flow variables.

We consider three types of possible steady states of an accretion disc: (i) stable low energy steady state (ii) energetic steady state (iii) unstable intermediate steady state. In the beginning of \S5 we will discusses the three types of steady states in terms of the steady state parameters of the accretion flow.  The following are the consistency checks of our model. (a) We developed the model with the assumption that the energetic initial steady state has enough free energy to induce state transition. Therefore if we choose the flow parameters such that the initial state is a stable low energy steady state then the system should not show any state transition. (b) By definition the intermediate state is hydrodynamicaly unstable. Therefore the hydrodynamic conditions of the model should also allow the system to fall back to more stable initial state. Also from the observational phenomenology we know that during aperiodic state transitions the system returns to the initial spectral state. Thus transition to an intermediate state should be reversible. 
(c) In the following sections \S 3 and 4, hydrodynamic conditions for state transitions are derived w.r.t a background disc obeying standard accretion disc equations. Therefore any plausible state transition should be consistent with Eqn.\ref{apprx}.

\section{Self-similar linear first order solutions}
The first order scale height ($h_{1}$) is a nonlinear function of first order flow variables. If the transition will not cause any inflated region in disc, then the ratio of the first order scale height to the zeroth order scale height ($h_{10}$) is close to unity, which means
\begin{equation}
\frac{h_{1}}{h_{0}} \equiv  h_{1,0}  =   \left(1+ \frac{c_{s1}^2 \rho_{0}}{c_{s0}^2 \rho_{1}} \right)^{\frac{1}{2}}  \approx \left(1+ \frac{c_{s1}^2 \rho_{0}}{2 c_{s0}^2 \rho_{1}} \right)
\label{lincnd}
\end{equation}
That is 
\begin{equation}
\sigma \ll 1  \;  \mbox{where}  \; \sigma  =  \frac{c_{s1}^2 \rho_{0}}{2 c_{s0}^2 \rho_{1}}
\label{lincnd1}
\end{equation}
Along with the above condition, for a self-similar first order solution, the first order fluid equations can be linearised. The self-similar linear first order solution is given by:
\begin{equation}
\rho_{1} = W r^{-\frac{3}{2}} \, \, , \,\, c_{s1}^2 = T r^{-1} \, \, , \, \, v_{1} = Y r^{-\frac{1}{2} }\, \, , \, \, \Omega_{1} = Z r^{-\frac{3}{2}}
\label{lssn}
\end{equation}
\noindent
such that the first order fluid equations become:
\begin{eqnarray}
&&W = A \left( \delta - \frac{Y}{C} - \frac{T}{2B} \right)\\[2ex]
&&\frac{5T}{2} + CY + 2DZ = 0\\[2ex]
&&\frac{3}{2} \alpha_{0}DT + \frac{\alpha _{0}}{\alpha} DY +\frac{3}{2} \left(\alpha_{0} - \alpha\right) BZ=\frac{3}{2} \left(\alpha - \alpha_{0}\right) BD\delta\\[2ex]
&&\frac{9}{4} f_{0}BD \left(\alpha + \alpha _{0}\right)Z - \frac{9}{4} f_{0} \alpha_{0} \frac{BD^2}{C}Y + \\[2ex] \nonumber
&&\left(  \frac{9}{4} f_{0} \alpha_{0} D^2 - \left( \frac{3\gamma}{2} - \frac{5}{2} \right) \frac{C}{\gamma-1}\right) T = \frac{9}{4} BD^2\left( \alpha f - \alpha_{0} f_{0} \right)\delta
\end{eqnarray}
It is evident from the above set of equations that solution does not exist if $\alpha = \alpha_{0}$ and $f = f_{0}$. Different  sets of plausible transitions and their corresponding values of $\sigma$ are listed in Table 1. It is seen that for transitions in advection dominated accretion disc the condition Eqn.$\ref{lincnd1}$ is valid. However in the case of  advective discs, for transitions from high viscous flow to low viscous flow the condition Eqn.$\ref{lincnd1}$ is not valid, which means that for such transitions one has to look for fully nonlinear solution. Moreover in the following  sections we find that the linearisation destroys some crucial properties of the disc. For example due to the original assumption Eqn.$\ref{lincnd1}$, linear solution does not allow an inflated state of the first order density $\rho_{1}$. Such an inflated state may source corona/outflow. Also in the following sections we see that the nonlinearity does not permit certain classes of transitions.

\begin{table}[t!]\caption{Parameters for linear transition with $\gamma=1.5$
}
\centerline{\begin{tabular}{lccccc}
\hline\hline\\[-2mm]
Set & $\alpha_{0}$ & $f_{0}$ & $\alpha$ & $f$ & $\sigma$\\
\hline
A1 & 0.01& 1    & 0.2  & 1 & 0.0053 \\
A2 & 0.01 & 1 & 0.1 & 1 &0.0013\\
A3 & 0.2& 1     & 0.01  & 1 &-0.00026597\\
A4 & 0.1 & 1& 0.01  & 1 &-0.00012606\\
\hline
B1 & 0.01& 1 & 0.2  & 0.5 &0.0084\\
B2 & 0.01 & 0.7 & 0.2  & 0.5 &0.0063\\
B3 & 0.2& 0.5     & 0.01 & 1 &0.3152\\
B4 & 0.2 & 0.5 & 0.01 & 0.7 & 3.0952\\
\hline
C1 & 0.01& 0.5  & 0.2  & 1 &0.0033\\
C2 & 0.01 & 0.5 & 0.2  & 0.7&0.0034\\
C3 & 0.2& 1     & 0.01 & 0.5&0.7075\\
C4 & 0.2 & 0.7 & 0.01 & 0.5 &0.4702\\
\hline
\label{tbl1}
\end{tabular}}
\end{table}

\section{Nonlinear first order equations}
It is evident from the above analysis that we cannot always rely on the self-similar linear first order solutions. Therefore we numerically solve the exact nonlinear first order conservation equations. Substituting the expression for $h_1$ (Eq.\ref{hexp1}) in first order continuity equation (Eq.\ref{fr1} given in Appendix~I), we can obtain a quadratic equation for first order density as:
\begin{equation}
a \rho_{1}^{2} \, + \, b \rho_{1} \, + \, c = \, 0 
\end{equation}
where
\begin{equation}
a = 1 ; \, b = \frac{A}{2B} r^{-1/2} c_{s1}^2  \,; \, c  =  -( A\delta r^{-3/2} - \frac{A}{C} r^{-1} v_{1} )^2
\end{equation}
such that
\begin{equation}
\rho_{1} \, = \, \frac{-b \pm \sqrt{b^{2} \, - \, 4ac }}{2a} 
\label{qdrsl}
\end{equation}
Thus there are two sets of nonlinear first order solutions. By substituting Eqn.\ref{qdrsl} in Eqns.\ref{fr2}-\ref{fr4} (see Appendix-I) we obtain the system of fully nonlinear first order equations. The effect of fluctuation in lower scale turbulence on the higher scale fluid transport is manifested as the variation in viscosity and energy advection. If the system has to maintain a nonlinear saturation then there should be a continuous energy cascade to the lower scale. In the  first order energy equation the advected energy of the zeroth order solution by the first order velocity is assumed to have an equipartition between the radial advection and cascade to the lower scale. The nonlinear first order equations are given by:
\begin{equation}
L_{1}\frac{\partial  v_{1} }{\partial r} + L_{2}\frac{\partial  c_{s1}^2 }{\partial r}  = L_{3}
\label{nln1}
\end{equation}
\begin{equation}
\alpha_{0} \left(\, L_{4} + L_{5}\frac{\partial  S }{\partial r} + L_{6}\frac{\partial  c_{s1}^2 }{\partial r} + L_{7}\frac{\partial  v_{1} }{\partial r} \right)\,  =  L_{8}
\label{nln2}
\end{equation}
\begin{equation}
L_{9}\frac{\partial  c_{s1}^2 }{\partial r}  + L_{10}\frac{\partial  v_{1} }{\partial r} +L_{11} = 0
\label{nln3}
\end{equation}
Expressions for $L_{n}$'s are given in the Appendix - II.

The total energy of a fluid particle is the sum of the radial kinetic energy, angular kinetic energy, the gravitational potential energy,  the internal energy and the pressure volume energy. Therefore the energy density of the accretion flow at any radius is given by
\begin{equation}
E \, = \, \rho \left(\, \frac{1}{2} v^{2} \, + \frac{1}{2} \, r^{2} \Omega^{2} \, - \, \frac{1}{r} \, + \, \frac{\gamma}{\gamma \, - \, 1} c_s^{2} \right)\,
\end{equation}
For an accretion flow which has undergone transition, the total energy density is the sum of the zeroth order component and the first order component. Thus we can define the first order energy density of the flow as
\begin{equation}
E_{1} \, = \, E (\rho \, , \, v \, , \, \Omega \, , \, c_s^{2})\, - \, E(\rho_{0} \, , \, v_{0} \, , \, \Omega_{0} \, , \, c^{2}_{s0})
\end{equation}
\subsection{Boundary conditions}
For numerical integration, we need to specify the values of first order flow variables at the outer boundary of the accretion flow domain. We choose the self-similar linear first order velocity and sound speed given by Eqn.\ref{lssn} as the initial values, which is a natural choice. 
In this work we choose the outer boundary of the sub-Keplerian accretion flow at $r=r_{o}=10^{3}$ and integrate up to $r=r_{i}=10$, below which the gravitational force varies considerably from its Newtonian value.

\section{Numerical results and discussion}
The hydrodynamic stability of an accretion disc is related to the choice of the combination of parameters of the flow \citep{NY94, NY95, MC05, RM10, RBM10}.
When the viscosity of the flow is low, the rate of dissipation of the angular momentum will also be low. As a consequence the residence time of the fluid particle in an orbit will be relatively large which increases the chance of ion-electron interaction and the efficiency of radiative cooling. Such a low viscous low advective phase is the stable low energy steady state of the disc.

If the viscosity of the flow is large then the rate of dissipation of angular momentum will also be large. As a consequence the fluid particle will fall to the inner orbit very quickly which decreases the chance of ion-electron interaction and efficiency of radiative cooling. In such high viscous high advective state, the flow contains large amount of energy. Even in this case the system will be relatively stable against various hydrodynamic instabilities because the matter will be swept away from an orbit before hydrodynamic instabilities become strong. The high viscous high advective phase or its limiting case of ADAF phase is an energetic state of the accretion disc.

Between these extreme limits, theoretically it is possible for the accretion disc to have a state with low viscosity and high advective factor. Since the viscosity is low the efficiency of angular momentum dissipation is low and the fluid particle could stay in an orbit for relatively large time. Since the advective factor is very high the efficiency of radiative cooling is low. Therefore non-radiative mechanisms such as convection will be triggered to shed energy \citep{IA99}. Such a low viscous high advective flow, in other words, the convection dominated accretion flow (CDAF) phase is an intermediate steady state of the accretion disc.
This state is not stable because of the inherent convective and diffusive instabilities,
as the effective buoyancy frequency becomes imaginary \citep{NY94}.
Outflows followed by spectral transitions are thought to be induced by convection which may
also give rise to quasi periodic oscillations. 

In this section we study the numerical solutions of the fully nonlinear first order conservation equations (Eqn.$\ref{nln1}$-$\ref{nln3}$) with respect to the self-similar background accretion disc (Eqn.\ref{solfn}). For a given state transition there could exist two branches of nonlinear first order solutions (see Eqn.\ref{qdrsl}). In the following subsections we analyse the two branches of solutions for different classes of transitions.  Throughout this section, we
choose $\dot{M}_{0} = 1$ for the mass accretion rate of the disc, without loss of generality as, both $\rho_0$ and $\rho_1$ scale proportionally to $\dot{M}_0$.
\subsection{Transitions in Advection Dominated Accretion Flows}
Underluminous sources are efficient in advecting viscously dissipated energy and are generally modelled as high viscous flows.
In this subsection we study the limiting case of ADAF. We explore the possibility of transitions of ADAF to a low viscous intermediate state, i.e., CDAF as discussed above, and study the response of the system to such transitions. For the initial ADAF state we make two choices for the viscosity:
(i) $\alpha_0=0.2$, and (ii) $\alpha_0=0.1$.
Transitions to a state of low viscosity with $\alpha=0.01$ is then explored.
For these transitions we have $f_0=f=1$ for the initial and intermediate advective factors. Having first explored such a forward transition, we then study the reverse transition from the intermediate state back to the high viscous state.

\subsubsection{Forward transition: from high to low viscous phase}

\begin{figure}
\centering
\includegraphics[width=0.90\columnwidth]{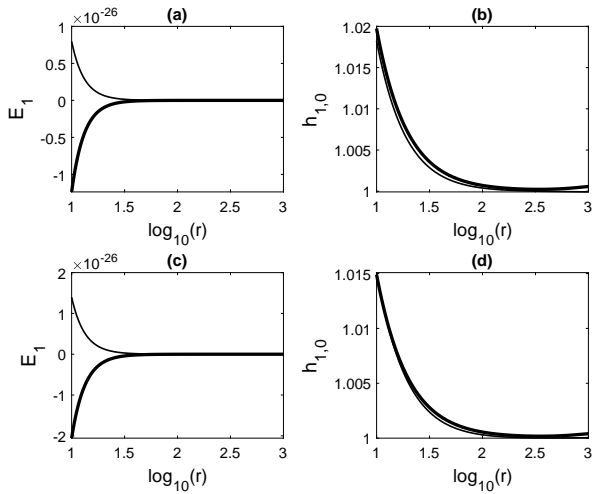}
\caption{
Radial dependencies of (a,c) $E_1$ and (b,d) $h_{1,0}$ in forward transition as
discussed in \S~5.1.1. Top (bottom) row: (i) $\alpha_{0}=0.2 \rightarrow  \alpha=0.01$
(ii. $\alpha_{0}=0.1 \rightarrow  \alpha=0.01$). Two curves in each panel correspond to the
two solution branches (see Eqn.\ref{qdrsl}); $\dot{M}_{0}=1$, $M=10$, $\epsilon=10^{-3}$,
$f_{0}, \,f = 1$ and $\gamma = 1.5$.
}
\label{nfig8} \end{figure}

Figure~\ref{nfig8} shows the two cases of state transition of the ADAF from high viscous phase to low viscous phase: case (i) $\alpha_{0}=0.2 \rightarrow  \alpha=0.01$, and
case (ii) $\alpha_{0}=0.1 \rightarrow  \alpha=0.01$.
In Fig.~\ref{nfig8} we show the first order energy density, $E_{1}$ (panel a),
and the ratio of first order scale height to zeroth order scale height, $h_{1,0}$
(panel b), corresponding to the two branches of solutions that appear for case (i).
One branch of solution has $E_{1}<0$ whereas $E_{1}>0$ for the other branch.
We find that both the solutions extend up to the inner region of the accretion disc,
with $h_{1,0} \simeq 1$. This means that the transition does not cause any inflated region
above the disc. The approximation as given in Eqn.\ref{lincnd} is therefore valid throughout
the disc and a self-similar linear first order solution can exist.
In bottom rows of Fig.~\ref{nfig8}, we display our findings for case (ii) where
behavior of $E_1$ and $h_{1,0}$ are shown in panels (c) and (d), respectively.
Qualitatively, the results are similar to case (i) as discussed above. In both these cases
we focus on the solution with $E_1<0$ which refers to the accretion solution.

\begin{figure}
\centering
\includegraphics[width=0.90\columnwidth]{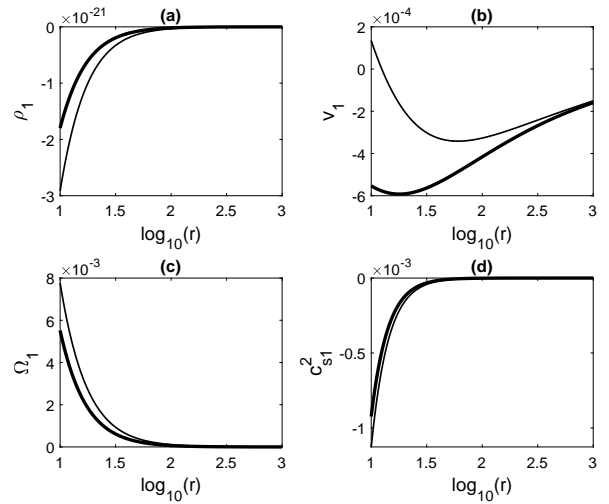}
\caption{
Radial dependencies of first order flow variables corresponding to $E_1<0$ solutions
of Fig.~\ref{nfig8}. Thick (thin) curve: (i) $\alpha_{0}=0.2 \rightarrow  \alpha=0.01$
(ii. $\alpha_{0}=0.1 \rightarrow  \alpha=0.01$).
}
\label{nfig9} \end{figure}

To further understand the accretion solution with $E_1<0$ that appears in this
CDAF state, we show in Fig.~$\ref{nfig9}$ the corresponding first order flow variables, namely,
$\rho_1$, $v_1$, $\Omega_1$, and $c_{s1}^2$. Thick and thin curves in each panel represent
cases (i) and (ii), respectively.
Nature of these variables reveals that the first order solutions of this CDAF state
tend to stabilize the system with $E_1<0$. For example, the first order radial velocity
$v_{1}$ is mostly negative which tends to push the matter radially inward, thus favouring
radial advection of viscous energy. Due to this, the matter falls more quickly causing a
decrease in effective density at each radius, i.e., $\rho_{1} <0$ (see panel a of
Fig.~$\ref{nfig9}$). As the first order flow favours stability of the disc the effective
temperature of the fluid decreases ($c^{2}_{s1}<0$),
and $\Omega_{1}$ turns out to be positive.
Below we explore the possibility of a reverse transition back to ADAF state from this
CDAF state.

\subsubsection{Reverse transition: from low to high viscous phase}

\begin{figure}
\centering
\includegraphics[width=0.90\columnwidth]{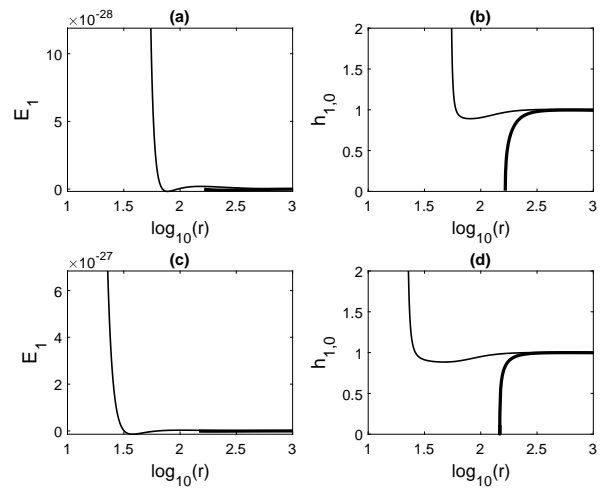}
\caption{
Same as Fig.~\ref{nfig8} but for the reverse transition as discussed in \S~5.1.2.
Top (bottom) row: (i) $\alpha_{0}=0.01 \rightarrow \alpha=0.2$
(ii. $\alpha_{0} = 0.01 \rightarrow \alpha = 0.1$).
}
\label{nfig10} \end{figure}

The ADAF phase is hydrodynamically more stable compared to the convectively unstable intermediate CDAF phase and therefore the system may eventually return to the initial state. Here we check if the hydrodynamic conditions of the model allow such a reverse transition.
We consider two cases to study such a reverse state transition from a low viscous
CDAF phase to a high viscous ADAF phase: case (i) $\alpha_{0}=0.01 \rightarrow \alpha=0.2$,
and case (ii) $\alpha_{0} = 0.01 \rightarrow \alpha = 0.1$.
As in the previous subsection, we show in Fig.$\ref{nfig10}$ the first order solutions
for $E_{1}$ and $h_{1,0}$, where the two curves in each panel correspond to the two
branches of solutions that are expected as noted above; top and bottom rows in 
Fig.$\ref{nfig10}$ correspond to cases (i) and (ii), respectively.

Let us first examine our findings in case (i).
The solution with $E_{1}<0$ terminates far away from the inner portion of the
accretion disc whereas the solution with $E_{1}>0$ extends close to the inner accretion
disc ($r\lesssim50$). Thus, only the first order solution with $E_{1}>0$ can communicate
the variation in matter, momentum and energy from the outer boundary to the inner portion of
the disc; focus on the thin curves in Fig.$\ref{nfig10}$(a) and (b). It is the physically acceptable solution where, at the inner
portion of the accretion disc, the first order scale height ($h_1$) is greater than the zeroth
order scale height ($h_0$) as $h_{1,0}>1$. We find that $h_{1,0}$ diverges at
$r\lesssim50$. Moreover, the first order density ($\rho_{1}$) is positive throughout the flow domain. This means that at the inner region of the accretion disc, the flow extends much
above the background accretion disc (the zeroth order solution). This low density hot plasma of mean density $\varepsilon \rho_1$,
and with positive energy ($E_1>0$) above the accretion disc could be interpreted as the formation of source of corona. If in this region the magnetic field of sufficient strength is coming out of the disc plane, then that could channelise this matter as outflow.

Thus the low viscous intermediate state is stabilised in such a state transition where
the gas expands at the inner portion of the accretion disc. This plausibly provides a source
for corona or outflow which is accompanied by a spontaneous transition of the system back to
the high viscous ADAF state. Once $h_{1,0}\gtrsim2$, the mean field theory of first order flow variables as discussed in \S2 is no longer valid and corresponding matter should be treated as
a separate flow domain. Bottom panels (c) and (d) of Fig. $\ref{nfig10}$ show $E_{1}$ and
$h_{1,0}$ for case (ii). Qualitatively the results are similar to case (i), except that
the first order flow extends a bit further inside the disc before diverging at
around $r\lesssim30$. 

\begin{figure}
\centering
\includegraphics[width=0.90\columnwidth]{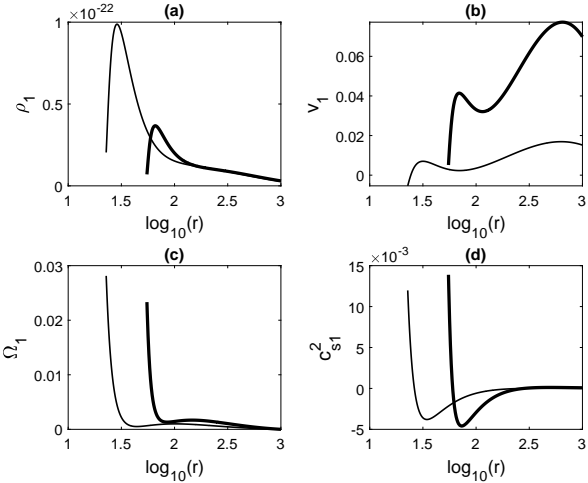}
\caption{
Radial dependencies of first order flow variables corresponding to $E_1>0$ solutions
of Fig.~\ref{nfig10}. Thick (thin) curve: (i) $\alpha_{0}=0.01 \rightarrow  \alpha=0.2$
(ii. $\alpha_{0}=0.01 \rightarrow  \alpha=0.1$).
}
\label{nfig11} \end{figure}

In Fig.~$\ref{nfig11}$ we show the first order flow variables corresponding to the
$E_{1}>0$ solution discussed above. The thick and thin curves there represent cases
(i) and (ii), respectively. In such state transitions, we find that $v_{1}>0$, and
therefore it reduces the effective inward radial velocity, causing a slow-down of the gas.
This leads to a pile up of matter at each radius causing an enhancement in the total
density, reflected by the fact that $\rho_{1}>0$.
At radii slightly larger than $\gtrsim50$ in case (i) (or $\gtrsim30$ for case ii)
we find that $h_{1,0}<1$, suggesting that the first order scale height $h_1$ decreases
there; see Figs.~$\ref{nfig10}$ and $\ref{nfig11}$. This causes a sharp increase in
the first order density solution $\rho_{1}$.
Since the effective inward radial velocity is decreased and the effective density is increased,
the ability to cool via ion-electron collision is enhanced, causing a decrease in
temperature of the gas at those outer portions of the disc where $c^{2}_{s1}<0$.
At smaller radii in the inner portion of the disc we have $h_{1,0}>1$. The gas expands there
and it causes a sharp decline in $\rho_{1}$, leading to a less efficient cooling
via ion-electron collision process. Therefore at the inner portion of the disc $c^{2}_{s1}$
is positive and it increases sharply as radius decreases. Furthermore, $\Omega_1$ is positive
and increases sharply as the flow diverges at inner portion of the disc.

\subsection{Transitions in high viscous advective flows}

Here we study the possible transitions of the flow which is in an initial high viscous phase of viscosity $\alpha_{0} = 0.2$, but not necessarily an ADAF. For the initial phase we choose a moderate value for the advective factor, $f_{0}=0.5$.
We first explore the transition to a low viscous intermediate phase of viscosity $\alpha = 0.01$,
and then study the reverse transition back to the high viscous phase.

\subsubsection{Forward transition: from high to low viscous phase}

In order to focus on some suitable choices for the advective factor $f$ in these
forward state transitions, we first study the nature of solutions that are obtained
for a wide range of $f$.
As noted earlier, we expect two solution branches in each case due to the two roots
as given in Eqn.~\ref{qdrsl}. We find that for very high values of $f$,
both the branches of the first order solutions exist, out of which only one corresponds
to the accretion solution. As $f$ decreases to a value of about 0.7, one branch of
the solution disappears, and at even lower values of $f$ there are no solutions.
Therefore when we restrict $f$ to take values $\gtrsim0.7$ we find that the hydrodynamic
conditions of the model allow state transitions of the accretion flow to a state
of low viscous high advective phase.

\begin{figure}[h]
\centering
\includegraphics[width=\columnwidth]{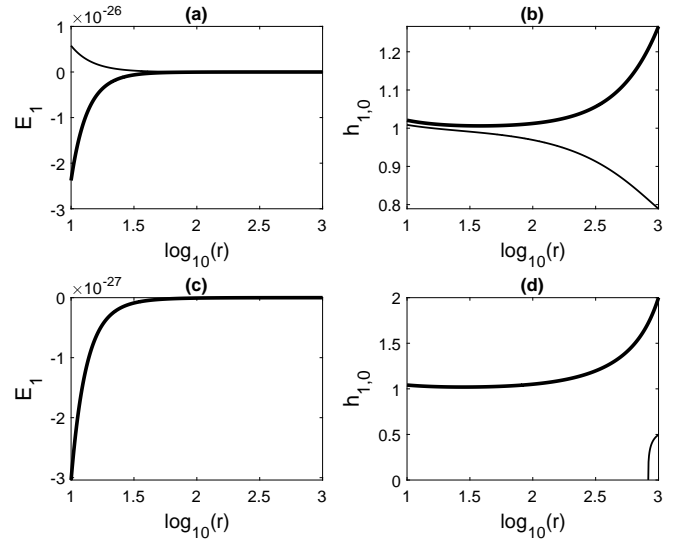}
\caption{
Similar to Fig.~\ref{nfig8} but for the forward transition as discussed in
\S~5.2.1 where $\alpha_{0}=0.2 \rightarrow \alpha =0.01$. Top (bottom) row:
(i) $f_0=0.5 \rightarrow f=1$ (ii. $f_0=0.5 \rightarrow f=0.7$).
}
\label{nfig1}
\end{figure}

Let the disc be initially in a state with $\alpha_0=0.2$ and $f_0=0.5$ which
undergoes a transition to a new intermediate state with $\alpha=0.01$. We consider
the following two cases where the advective factor changes in these transitions as:
case (i) $f_0=0.5 \rightarrow f=1$, and case (ii) $f_0=0.5 \rightarrow f=0.7$.
In Fig.~$\ref{nfig1}$ we show results from cases (i) and (ii) in top and bottom rows,
respectively, where the two curves in each panel correspond to the two solution
branches. We show again the radial dependence of $E_1$ and $h_{1,0}$, and discuss
first the solutions of case (i). One branch has $E_1<0$ (accretion solution; thick
curves) and the other has $E_1>0$. Both solutions extend up to the inner regions of
the disc. At outer boundary $h_{1,0}$ differs from unity. This is because of the
choice of the initial condition.
To find the exact numerical solution we choose self-similar linear solution as the initial
value; see \S~3. From set B3, B4 of Table~1 we see that the condition $\sigma\ll1$
(Eq.~\ref{lincnd1}) is not valid for transitions of general advective flows, when the transition
is from a high viscous flow to a low viscous flow. Therefore $h_{1,0}$ is
different from unity in the beginning, but when the nonlinearity picks up, $h_{1,0}$ quickly becomes
close to unity. The first order variables do not cause any inflated region, i.e., as $r$ decreases
we find $h_{1,0}\sim1$. 
In the case (ii) we see that only one branch of solution exists which has $E_{1}<0$ (the
accretion solution); see bottom row of Fig.~\ref{nfig1}.
The other branch of solution terminates far away from the inner boundary.
The variations of the first order flow variables are qualitatively similar to the
corresponding case of ADAF discussed in \S 5.1.1 and hence are not shown here.

\subsubsection{Reverse transition: from low to high viscous phase}

\begin{figure}
\centering
\includegraphics[width=0.90\columnwidth]{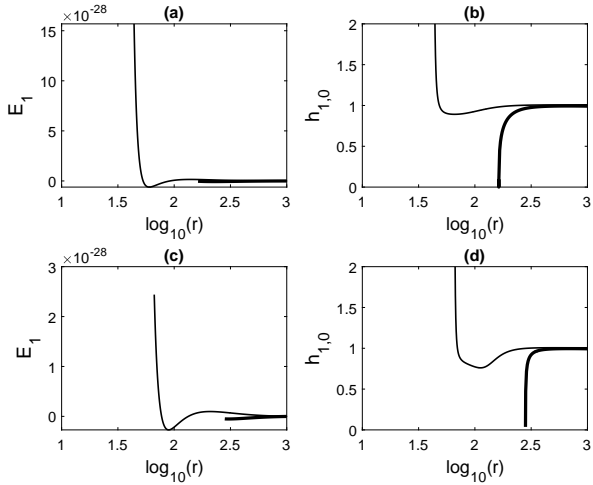}
\caption{
Similar to Fig.~\ref{nfig8} but for the reverse transition as discussed in
\S~5.2.2, where $\alpha_{0}=0.01 \rightarrow \alpha =0.2$. Top (bottom) row:
(i) $f_0=1 \rightarrow f=0.5$ (ii. $f_0=0.7 \rightarrow f=0.5$).
}
\label{nfig3} 
\end{figure}

\begin{figure}
\centering
\includegraphics[width=0.90\columnwidth]{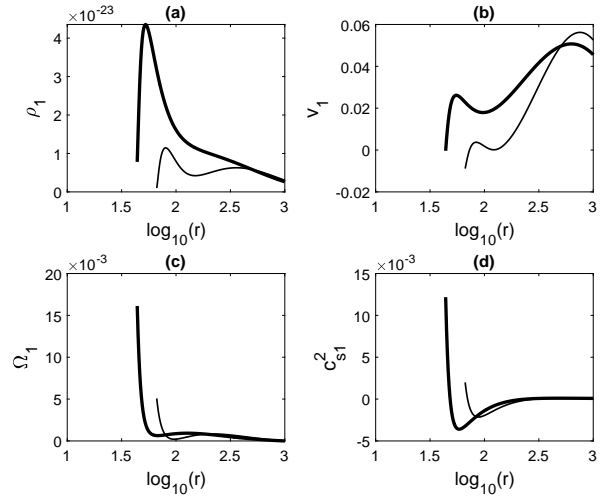}
\caption{
Radial dependencies of first order flow variables corresponding to $E_1>0$ solutions
of Fig.~\ref{nfig3}, where $\alpha_{0}=0.01 \rightarrow \alpha =0.2$.
Thick (thin) curve: (i) $f_0=1 \rightarrow f=0.5$
(ii. $f_0=0.7 \rightarrow f=0.5$).
}
\label{nfig4} 
\end{figure}

As we saw in \S~5.1.2, the system may eventually return back to the initial
state of high viscous advective phase which is hydrodynamically more stable compared
to the intermediate low viscous high advective or CDAF phase. Let us
check if the hydrodynamic conditions of the model allow such a reverse transition
where $\alpha_{0}= 0.01 \rightarrow \alpha = 0.2$, for which we consider the
following two cases: case (i) $f_{0}=1 \rightarrow f=0.5$, and case (ii)
$f_{0}=0.7 \rightarrow f=0.5$. Both these cases turn out to be qualitatively
similar, and are shown in Fig.$\ref{nfig3}$. Let us focus therefore only on
panels (a) and (b) representing the case (i). 
The solution with $E_1<0$ terminates far away from the the inner boundary of the accretion disc, whereas the solution with 
$E_1>0$ extends close to the inner portion of the accretion disc. Therefore the solution with $E_1>0$ is the physically acceptable solution.
As in \S~5.1.2, we find again that
the first order solutions $E_1$ and $h_{1,0}$ diverge at $r\lesssim50$,
suggesting thus that the flow extends much above the background disc in
the inner regions as the scale height $h_1>h_0$ there. 
Reverse transition in these cases too are able to provide a source for corona or outflow.
First order flow variables ($\rho_1, v_1, \Omega_1, c_{s1}^2$) corresponding
to the $E_1>0$ branch are shown in Fig.\ref{nfig4}. These behave identically to those
shown in Fig.~$\ref{nfig11}$; we therefore refer the reader to \S~5.1.2 for a discussion
on the nature of these solutions.

\subsection{Transitions in low viscous advective flow}

Low viscous, radiatively efficient phase of accretion flow is generally associated with a low energy stable steady state. Here we choose such an initial state of viscosity with $\alpha_{0}=0.01$, and moderate advective factor $f_{0}=0.5$. We first explore the possibility of a transition to an ADAF ($f=1$) phase or a highly advective phase with an advective factor,
e.g., $f=0.7$, and viscosity $\alpha=0.2$. Having studied such a forward transition, we then explore the possibility of a reverse transition back to the low energy steady state which has
small viscosity.

\subsubsection{Forbidden transition}

\begin{figure}
\centering
\includegraphics[width=1\columnwidth]{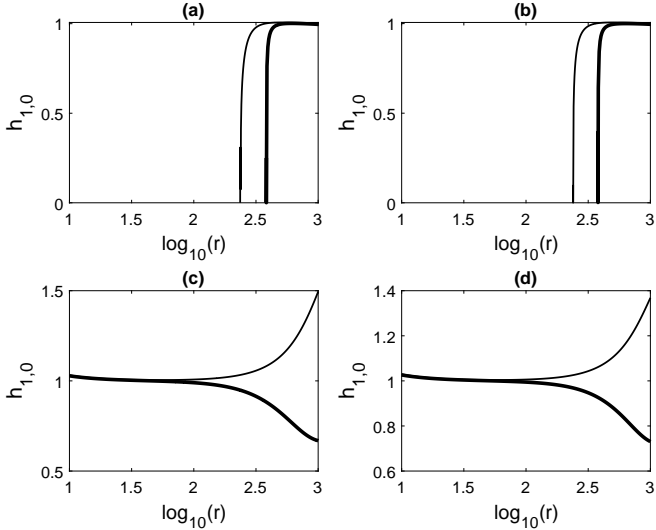}
\caption{
Radial dependence of $h_{1,0}$, the ratio of first order to zeroth order scale
heights; $\gamma = 1.5$. Two curves in each panel correspond to the two solution branches
(see Eqn.\ref{qdrsl}). Top (forbidden): $\alpha_{0} = 0.01  \rightarrow \alpha = 0.2$
where (a) $f_{0} = 0.5 \rightarrow f = 1$, and (b) $f_{0} = 0.5 \rightarrow f = 0.7$.
Bottom (allowed): $\alpha_{0} = 0.2  \rightarrow \alpha = 0.01$
where (c) $f_{0} = 1 \rightarrow f = 0.5$, and (d) $f_{0} = 0.7 \rightarrow f = 0.5$.
}
\label{nfig5} \end{figure}

Upper row of Fig.$\ref{nfig5}$ shows the radial dependence of $h_{1,0}$ in the state
transition where $\alpha_{0}= 0.01\rightarrow \alpha = 0.2$. Panels (a) and (b)
correspond, respectively, to case (i) $f_{0} = 0.5 \rightarrow f=1$ and
case (ii) $f_{0}= 0.5 \rightarrow f=0.7$.
In both the cases the two branches of solutions terminate far away from the inner region of the disc. Therefore none of the first order solutions can carry the change in matter, momentum and energy from the outer boundary to the inner portion of the accretion disc. Thus no physically acceptable first order solution exists, or in other words, this particular class of transition is not possible.

The assumption of the model developed in \S2 is that the intrinsic fluctuations in flow parameters can extract energy from an energetic background flow, and they can
grow and saturate to new steady state values. Since a low viscous and low/moderately advective phase is a low energetic initial state of the accretion disc, the background accretion flow cannot provide enough free energy to drive the transition. This explains the non-existence of the first order solution.

\subsubsection{Allowed transition}
Here we study if the accretion flow in an energetic state such as ADAF/high viscous highly advective phase 
can spontaneously relax to a low energetic state such as low viscous low/moderate advective phase. 
Transitions exist for moderate values of $f$. As $f\lesssim0.3$ the transition ceases to happen. We present the reverse transition of the cases discussed in \S 5.3.1. Lower row of Fig.$\ref{nfig5}$ shows the radial dependence of $h_{1,0}$ in the state
transition where $\alpha_{0}= 0.2\rightarrow \alpha = 0.01$. Panels (c) and (d)
correspond, respectively, to case (i) $f_{0} = 1 \rightarrow f=0.5$ and
case (ii) $f_{0}= 0.7 \rightarrow f=0.5$. Similar to the cases discussed in \S 5.1.1 and \S 5.2.1, here also two branches of solutions exists out of which the one with $E_1<0$ is the accretion solution. The solutions extend up to the inner boundary of the accretion disc. Similar to the cases discussed in \S 5.2.1 here also the condition $\sigma<<1$ is not valid (see set C3, C4 of Table 1). Hence the initial value of $h_{1,0}$ is different from unity. Once the nonlinearity picks up, $h_{1,0}$ becomes close to unity. The transition does not cause any inflated region ie; as the radius decreases, $h_{1,0}\sim1$.

\subsection{Self-consistency of the model}

\begin{figure}
\centering
\includegraphics[width=1\columnwidth]{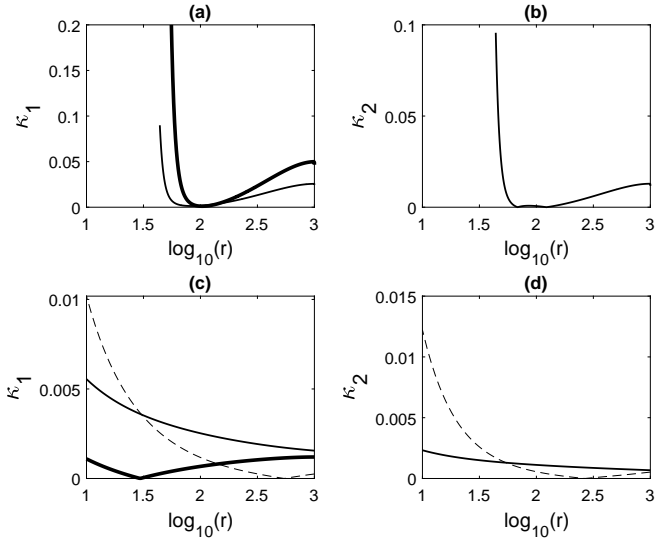}
\caption{Variation of $\kappa_{1}$ and $\kappa_{2}$ for different classes of transitions;
$\epsilon = 10^{-3}$. (a) and (b) are for transitions $\alpha_{0}= 0.01 \rightarrow \alpha =0.2$, where thick
and thin curves correspond to $f_{0}, f=1 $ and $f_{0}= 1$, $f= 0.5 $, respectively.
Panels (c) and (d) are for transitions $\alpha_{0} = 0.2 \rightarrow \alpha = 0.01$,
where thick and thin curves correspond to $f_{0}, f = 1 $ and $f_{0} = 0.5$, $f = 1 $,
respectively; dashed curves are for $f_{0} = 1$ and $f= 0.5$.
}
\label{nfig12} \end{figure}

We now test whether the results of our numerical analysis are self-consistent; see the
end of \S 2 where three consistency checked are mentioned.

(a) Our model assumes that the initial state of the system is energetic such that enough
free energy is available for state transition. A low viscous and low advective phase of
the accretion flow is a low energetic steady state which is stable.
In \S 5.3.1, we have seen that such a state does not show state transition. This result is
thus consistent with the assumption. Such a low energetic state can acquire energy from
environment by finite variation in mass accretion rate ($\varepsilon \, \gtrsim \, 1$),
which is beyond the scope of the present model. Also it is observed from the numerical
results presented in \S 5.1 and 5.2 that, in general, hydrodynamic conditions allow state
transitions only for $f_{0} , \, f  \gtrsim 0.5$. 

(b) Intermediate hydrodynamic steady state was deemed unstable such that the system
is expected to return to the initial state. In \S 5.1 and \S 5.2 we find that whenever
there is a state transition to an intermediate steady state, our model permits the
corresponding reverse transition. This may help explain, to some extent, the
observations of aperiodic variabilities.

(c) The zeroth order solution is chosen as a self-similar accretion disc obeying the
standard conservation equations. This choice is possible due to the approximation
given in Eqn.~$\ref{apprx}$. In Fig.\ref{nfig12} we have shown $\kappa_1$ and $\kappa_2$ for the first cases, (i), of the transitions discussed in \S5. Fig.\ref{nfig12}(a) shows $\kappa_1$ for the reverse
transitions discussed in \S5.1.2 and \S5.2.2. Equation~\ref{apprx} is valid till the inner
portion of the accretion disc. At $r \sim 50$ the first order scale height
(or equivalently, $h_{1,0}$) diverges, and the matter contained in the first order density
should be treated as a separate flow domain with an outflow rate $\varepsilon \dot{M}_0$.
Therefore the mean-field theory discussed in this work is no longer valid below this radius,
which explains the breakdown of the Eqn.\ref{apprx}. For the reverse transition discussed in \S 5.1.2, $\kappa_2$ is zero by definition;
$\kappa_2$ being proportional to $|f-f_0|$ vanishes as $f=f_0=1$.
For the reverse transition as discussed in \S 5.2.2, the corresponding $\kappa_2$ is shown
in Fig.\ref{nfig12}(b) which has similar behavior as $\kappa_1$.
Figure~\ref{nfig12}(c) and (d) shows $\kappa_1$ and $\kappa_2$, respectively, for the forward transitions as well as the transition discussed in \S 5.3.2. All of them obey the condition Eqn.\ref{apprx}. 


\section{Summary and Conclusions}

\begin{figure}
\centering
\includegraphics[width=0.7\columnwidth]{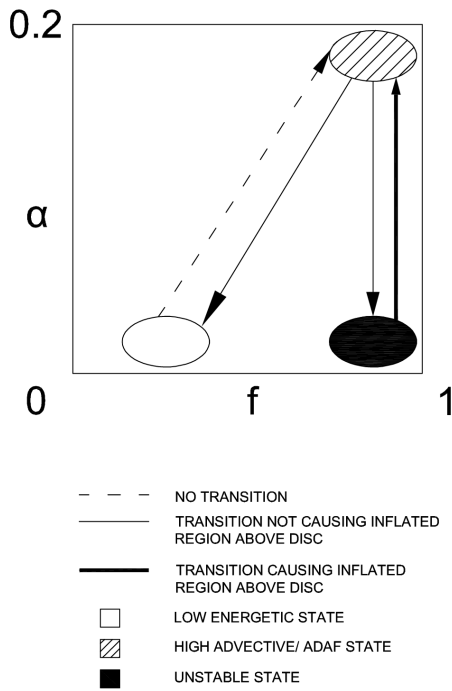}
\caption{Schematic diagram of the state transitions 
}
\label{nfigsch} \end{figure}

In this work we have presented a phenomenological model as an attempt to
understand the origin of different spectral states that are observed from
accretion discs around compact objects.
Different spectral states of an accretion system are modelled as a consequence of
transition among hydrodynamic steady states with corresponding
flow parameters, i.e., viscosity ($\alpha$) and advective factor ($f$).
A self-consistent choice of viscosity and efficiency of energy advection produces
transitions among different phases of accretion flow, such as radiatively
efficient accretion flow, ADAF, CDAF etc.
In Fig.\ref{nfigsch} we show a schematic diagram representing various state transitions
discussed in this work. The observational phenomenology
included in the present model are given as follows:
\begin{enumerate}
    \item [(a)] The system stays in different spectral states of aperiodic
    variabilities for considerable amount of time, e.g., Cyg X-1, LMC X-1,
    LMC X-3, GX 339-4 \citep{TH72, WNP01, GMM06}. Therefore we model the
    underlying accretion flow in different spectral states as having different
    hydrodynamic steady states.
    \item [(b)] Observed aperiodic variabilities are often followed by loss of
    matter from the disc which can source the corona/outflow. Therefore, the flow
    variables in this model are designed to incorporate a first order component,
    which, under suitable state transition could represent loss of matter.
    \item [(c)] Since the system often returns to the initial state, an intermediate
    steady state must have information about the initial state of the system.
    In this model, this information is contained in the first order variables of
    the intermediate state.
\end{enumerate}

\noindent

Aperiodic variabilities such as transient behaviour of X-ray binaries are transitions among spectral states like High-Soft State, Low-Hard State and other intermediate states \citep{BJ10, MV97, MRR09}. Often the excited states are followed by outflow \citep{MTG18}. In \S5.1.1 and \S5.2.1 we found that the model allows the state transition of the disc from ADAF or high viscous high advective phase to a low viscous high advective/convection dominated  intermediate phase. In \S5.1.2 and \S5.2.2 we have shown that, close to the inner boundary ($r\sim50$), the intermediate state will inflate the incoming first order component of matter and simultaneously return to ADAF or high viscous high advective phase. ADAF is generally associated with the Low-Hard State of the accretion system. The coexistence of ADAF and a hot low dense corona of
mean density $\varepsilon \rho_1$ and energy density $E_{1} >0$ is particularly important because such an explicit correlation is useful to model the scattering of the soft X-rays from disc by corona \citep{GRV79, EWM94, NP94, KB16}. The condensation of corona due to gravitational capture by the very inner portion of the accretion disc is another interesting related phenomenon \citep{MLQ17}. The accretion disc and the corona/outflow are though of as
two separate flow domains. In the present model the source of plausible corona/outflow is a natural outcome of the state transition. In the model, physical properties of this source is explicitly determined by the parameters of the background accretion flow. Such state transitions are possible for a range of values of advective factor such that $f \, \geq \, 0.5$ and $f_{0} \, \geq \, f$.
We envisage that if strong enough vertical magnetic fields are present, then it could
channelise the matter causing an outflow.
 
In \S5.3.1 we have shown that a low viscous moderate advective accretion disc cannot
make a transition to a high viscous high advective/ADAF state by the intrinsic tendencies of the flow. Whereas in \S5.3.2 we have shown the possibility of spontaneous transition from
a high viscous high advective/ADAF state to a low viscous moderate advective state.
Underluminous AGN's and quasars such as SgrA* are modelled as high viscous highly advective/ADAF accretion flows. Our model suggests that such energetic states are not natural and they will
make a transition to energetically more stable low viscous low/moderate advective state. The accretion disc could stay in such a high viscous high advective/ADAF state only if the system has a continuous supply of external energy from the outer disc/environment. 
For this to happen the environment in which the accretion system is situated should be
energetically very active; systems such as SgrA* are known to be situated in
active environments which may continuously supply the required energy \citep{CNW15, CC20}. 
The X-ray binary Cyg X-1 is found to be in Low Hard State for the most of the
time \citep{BJ10, WNP06}. 
We argue that the so called failed relaxation to low energy state could be understood as a
series of transitions among excited states propelled by continuous external energy input to
the system \citep{XY16}.
Spectral hardening may be understood to be a result of the transitions among
energetic steady states of the disc and the transitions induced corona/outflow.

\section*{Acknowledgments}
We thank Harsha Raichur for critical comments on an earlier version
of the manuscript.
This work is partly supported by facilities provided by the Department of Science and Technology, Government of India under the FIST programme. 
AA thanks the Council for Scientific and Industrial Research, Government of India, for the research fellowship. SRR thanks IUCAA, Pune for the Visiting Associateship Programme.

\section*{Appendix - I}
Substituting expressions of the flow variables (Eqn.\ref{vaf2}-\ref{vaf4}), vertically integrated stress tensor (Eqn.\ref{strsexp}), vertically integrated energy advection (Eqn.\ref{enadvexp}) and total mass accretion rate (Eqn.\ref{exp2}) in Eqn.\ref{ozr1} to \ref{ozr4}, and equating the first power of $\varepsilon$ we obtain the first order conservation equations:
\begin{equation}
4 \pi r \rho_{0}  h_{0} v_{1} + 4 \pi  r\rho_{1} h_{1} v_{0} = \dot{M_{1}}
\label{fr1}
\end{equation}

\begin{equation}
\rho_{0} v_{0} \frac{\partial  v_{1} }{\partial r} + \rho_{1} \phi^{\frac{1}{2}} v_{0} \frac{\partial  v_{0} }{\partial r} +\rho_{0} v_{1} \frac{\partial  v_{0} }{\partial r}  - \rho_{1}\phi^{\frac{1}{2}}\Omega_{0}^2r - 2\rho_{0}\Omega_{0}\Omega_{1} r  = - \frac{\partial  (\rho_{0} c_{s1}^2) }{\partial r}- \phi^{\frac{1}{2}} \frac{\partial  (\rho_{1}  c_{s0}^2) }{\partial r} - \frac {\rho_{1}\phi^{\frac{1}{2}}} {r^2}
\label{fr2}
\end{equation}

\begin{equation}
\rho_{0}h_{0} v_{1} \frac{\partial  (\Omega_{0} r^2) }{\partial r} +\rho_{1}h_{1} v_{0} \frac{\partial  (\Omega_{0} r^2) }{\partial r}+\rho_{0} h_{0}v_{0} \frac{\partial  (\Omega_{1} r^2) }{\partial r}  = \frac {1}{r} \frac{\partial  \left(r^2  \alpha_{0} D_{1} \, + \, \alpha_{1} D_{2} \right)}{\partial r}
\label{fr3}
\end{equation}
\begin{equation}
\frac{v_{0}}{(\gamma-1)} \left[  h_{0}\frac{\partial  (\rho_{0}c_{s1}^2) }{\partial r} +h_{1}\frac{\partial  (\rho_{1} c_{s0}^2) }{\partial r} - \gamma  c_{s1}^2 h_{0} \frac{\partial  \rho_{0} }{\partial r} - \gamma  c_{s0}^2  h_{1} \frac{\partial  \rho_{1} }{\partial r}\right]  + \frac{v_{1} h_{0}}{(\gamma-1)} \left[  \frac{\partial  (\rho_{0} c_{s0}^2) }{\partial r} -  \gamma  c_{s0}^2 \frac{\partial  \rho_{0} }{\partial r}\right]  
= f_{0} F_{1} +  f_{1} F_{2}
\label{fr4}
\end{equation}
\section*{Appendix - II}
The terms in the fully nonlinear first order equations with respect to a self-similar background disc (Eqn.\ref{nln1}-\ref{nln3}) are:
\begin{equation}
L_{1} = ACr^{-2} + Br^{-1}\phi^\frac{1}{2}k_{6}
\;;\quad
L_{2} = Ar^\frac{-3}{2} + Br^{-1}\phi^\frac{1}{2}k_{5}
\end{equation}
\begin{equation}
L_{3} =\left[ \frac{1}{2} ACr^{-3} v_{1} + 2AD r^{-2}\Omega_{1}+\frac{3}{2} Ar^\frac{-5}{2} cs_{1}^2 \right] - \rho_{1}\phi^{\frac{1}{2 }}\left[ 1- \frac{C^2}{2} -D^2 - B_{f}\right] r^{-2}- Br^{-1}\phi^\frac{1}{2} k_{4}
\end{equation}

 \begin{equation}
 L_{4} = 3ABrS - \frac{9}{4}ADr^\frac{-1}{2} cs_{1}^2 - 3BD\rho_{1} \phi ^ {\frac{1}{2}}- \frac{3}{2}BD r \left[\frac{\phi +1}{2\phi^ \frac{1}{2}} \right]k_{4} - \frac{3}{8}BD\rho_{1} \left[\frac{\phi -1}{\phi^ \frac{1}{2}} \right]
 \end{equation}
 
\begin{equation}
L_{5} = ABr^2
\;;\quad
L_{6} = - \left(\frac{3}{2}ADr^\frac{1}{2} +  \frac{3}{2}BD r \left[\frac{\phi +1}{2\phi^ \frac{1}{2}}\right] k_{5} + \frac{3}{4}BD r \left[ \frac{\phi -1}{\phi^ \frac{1}{2}} \right]\frac{\rho_{1}}{cs_{1}^2}\right) 
\;;\quad
L _{7}= -\frac{3}{2}BD r \left[\frac{\phi +1}{2\phi^ \frac{1}{2}}\right] k_{6} 
\end{equation}
\begin{equation}
L_{8} = 2AC\Omega_{1} + ACrS + \frac{1}{2} AD r^ {-1} v_{1} + \frac{1}{2} CD \rho_{1} \phi^\frac{1}{2} 
\;;\quad
L_{9} = \frac{1}{\gamma-1} AC r^{-1} - BC \phi^\frac{1}{2} r^{\frac{-1}{2}} k_{5} 
\;;\quad
L_{10} = - BC \phi^\frac{1}{2} r^{\frac{-1}{2}} k_{6}
\end{equation}
\begin{equation}
L_{11} = L_{111} + L_{112} + L_{113} + L_{114} + L_{115}
\;;\quad
 L_{111} = \frac{3}{2} f_{0} ABD r^{\frac{-1}{2}} ( \alpha + \alpha_{0} ) S
 \;;\quad
 L_{112} = \left(\frac{3\gamma-5}{4(\gamma-1)}\right) AB r^{\frac{-5}{2}} v_{1}
 \end{equation} 
 \begin{equation}
 L_{113} = - \left( \frac{9}{4}  AD^2f_{0} \alpha_{0} - \frac{3}{2}AC \right)r^{-2 }cs_{1}^2
 \;;\quad
 L_{114} = - \phi^\frac{1}{2} \rho_{1} r^{\frac{-3}{2}} \left( \frac{1}{\gamma-1} BC + \frac{9}{4}  BD^2 f_{0} \alpha_{0} \right)
 \;;\quad
 L_{115} =  - BC \phi^\frac{1}{2} r^{\frac{-1}{2}} k_{4}
 \end{equation}
\begin{equation}
\theta = \frac{A^2}{B^2} r^{-1} (cs_{1}^2)^2 +4 [ A\delta r^ \frac{-3}{2} - \frac{A}{C} r^{-1} v_{1}]^2
\;;\quad
k = 4 [ A\delta r^ \frac{-3}{2} - \frac{A}{C} r^{-1} v_{1}]
\;;\quad
k_{1} = 2k \left[ \frac{A}{C} v_{1}r^{-2} - \frac{3}{2} A\delta r^\frac{-5}{2}\right]  - \frac{A^2}{B^2}\left(\frac{cs_{1}^2}{r}\right)^2 
\end{equation}
\begin{equation}
k_{2} = 2 \frac{A^2}{B^2} r^{-1} cs_{1}^2
\;;\quad
k_{3} = - \frac{2Ak}{C} r^{-1} 
\;;\quad
k_{4} = \frac{A}{4B} r ^\frac{-3}{2} cs_{1}^2 \pm \frac{1}{4} \theta^\frac{-1}{2} k_{1}
\;;\quad
k_{5} = -\frac{A}{2B} r ^\frac{-1}{2} \pm \frac{1}{4} \theta^\frac{-1}{2} k_{2}
\;;\quad
k_{6} = \pm \frac{1}{4} \theta^\frac{-1}{2} k_{3}
\end{equation}
where
\begin{equation}
\frac{\partial h_{1}}{\partial r} = \frac{h_{0}\phi^\frac{1}{2} }{r} + \frac{h_{0}}{2} \left( \frac{{\phi}-1}{\sqrt{\phi}} \right)  \left[ \frac{1}{c_{s1}^2} 
\frac{\partial c_{s1}^2}{\partial r} - \frac{1}{\rho_{1}} \frac{\partial\rho_{1}}{\partial r} - \frac{1}{2r} \right] 
\label{nln4} \, \, \mbox{and} \, \, S=\frac{\partial \Omega_{1}}{\partial r}
\end{equation}

\end{document}